\documentstyle[12pt]{article}
\begin{document}
\thispagestyle{empty}
\baselineskip 24pt
\begin{flushleft}
\end{flushleft}  
\begin{center} 
{\large TM-SURFACE WAVES ALONG THE BOUNDARY BETWEEN TWO NONLINEAR 
ANISOTROPIC DIELECTRICS} \\ Yu.\ P.\ Rybakov$^{\star}$, B.\ 
Saha$^{\dagger}$ \\ $^{\star}$ Russian Peoples' Friendship University, 
Moscow \\
e-mail:rybakov@udn.msk.su \\
$^{\dagger}$ Joint Institute for Nuclear Research, Dubna\\
e-mail:saha@theor.jinrc.dubna.su
\end {center} \vskip 3mm 
{\it It is shown that the Maxwell's equations for surface electromagnetic 
TM waves, propagating along the plane boundary between two 
nonlinear dielectrics with arbitrary diagonal tensor of dielectric 
permittivity, depending on $|{\bf E}|$, can be integrated in quadratures.}
\newpage
Surface electromagnetic waves, propagating along the boundary between 
two media with different optical properties were widely studied 
[1-7].  Nonetheless the integration of the corresponding equations 
presents difficulties because of their complicated nonlinear form. 
We will consider surface TM waves for the case of two 
dielectric media with the diagonal dielectric permittivity tensor 
$\varepsilon_{i\, k}\,= \, \mbox{diag}(\varepsilon_x,\, \varepsilon_y,\, 
\varepsilon_z)$, therewith $\varepsilon_x,\, \varepsilon_y,\, 
\varepsilon_z$ are considered as arbitrary real functions of electric 
field strength ${\bf E}$.

Suppose the plane $x\,=\,0$ to be that of separation and plane 
electromagnetic wave with frequency $\omega$ and wave vector ${\bf k}$ to 
propagate along z-axis. We will consider TM wave, assuming
$${\bf H}\,=\,(0,\,H,\,0); \quad {\bf E}\,=\,(E_x,\,0,\,E_z)\,.$$
Putting $\partial_t\,=\,-v\,\partial_z$, where 
$v\,=\,\omega/k\,=\,\beta\,c$ is the wave velocity, we write the system 
of Maxwell's equations in each medium:  
\begin{equation}
\partial_x\,(\varepsilon_x\,E_x)\,+\,\partial_z\,(\varepsilon_z\,E_z)\, 
=\,0\,,
\end{equation} 
\begin{equation}
\partial_x\,(H\,-\,\beta\,\varepsilon_x\,E_x)\,=\,0\,,
\end{equation}
\begin{equation}
\partial_z\,E_x\,-\,\partial_x\,E_z\,=\,\beta\,\partial_z\,H\,.
\end{equation}
Inasmuch at $x\,=\,\pm\,\infty$ the field is supposed to vanish, 
from (2) we will find 
\begin{equation}
H\,=\,\beta\,\varepsilon_x\,E_x\,.
\end{equation} 
Assuming that $\partial_z\,=\,ik, \quad E_x\,=\,iA,\quad E_z\,=\,B$, 
where $A$ and $B$ are real functions, from (1), (3) and (4) we deduce 
\begin{equation}
(\varepsilon_x\,A)^{\prime}\,+\,k\,\varepsilon_z\,B\,=\,0\,,
\end{equation}
\begin{equation}
B^{\prime}\,=\,k\,A(\varepsilon_x\,\beta^2\,-\,1)\,,
\end{equation}
where prime denotes differentiation with respect to $x$. Taking the 
functions $\varepsilon_x(A^2\,+\,B^2)$ and $\varepsilon_z(A^2\,+\,B^2)$ as 
given ones, we will admit that $|{\bf E}|^2$ and $\varepsilon_z$ can be 
expressed as some functions of $\varepsilon_x$, i.e.  
\begin{equation}
|{\bf E}|^2\,=\,A^2\,+\,B^2\,=\,I(\varepsilon_x),\qquad \varepsilon_z(|{\bf 
E}|^2)\,=\,K(\varepsilon_x)\,.
\end{equation} 
Denoting $\varepsilon_x\,A\,=\,f$, from (5) and (7) we will find 
\begin{equation}
I(\varepsilon_x)\,=\,\frac{f^2}{\varepsilon_{x}^{2}}\,+ 
\,\frac{{f^{\prime}}^2}{k^2\,K^2(\varepsilon_x)}\,,
\end{equation} 
where, according to (5) 
\begin{equation}
B\,=\,-\,\frac{f^{\prime}}{k\,K(\varepsilon_x)}\,.
\end{equation} 
Differentiating the equation (8) in view of (6), one gets
\begin{equation}
\frac{d I}{d \varepsilon_x}\,\varepsilon_{x}^{\prime} \,=\, (f^2)^{\prime} 
\biggl(\frac{1}{\varepsilon_{x}^{2}}\,-\, \frac{\beta^2}{K}\,+\, 
\frac{1}{\varepsilon_x\,K}\biggr)\,-\,\frac{2\,f^2}{\varepsilon_{x}^{3}}\, 
\varepsilon_{x}^{\prime}\,.
\end{equation}
The equation (10) admits the integrating factor $Y(\varepsilon_x)$ 
(see Appendix 1):  \begin{equation} 
\mbox{ln}\,Y(\varepsilon_x)\,=\,\int\limits_{}^{\varepsilon_x}\,\frac{d 
K}{X\,K^2} \biggl(\frac{1}{\varepsilon_x}\,-\,\beta^2\biggr)\,+\, 
\int\limits_{}^{\varepsilon_x}\,\frac{d 
\varepsilon_x}{\varepsilon_{x}^{2}\,K\,X}\,,
\end{equation} 
where 
\begin{equation}
X(\varepsilon_x)\,=\,\frac{1}{\varepsilon_{x}^{2}}\,-\, \frac{\beta^2}{K}\, +\, 
\frac{1}{\varepsilon_x\,K}\,.
\end{equation} 
Given $\varepsilon_x(0)$, the value of 
permittivity $\varepsilon_x$ at ${\bf E}\,=\,0$, i.e. at $x\,=\, \pm 
\infty$, from (10) we deduce the quadrature 
\begin{equation}
\int\limits_{\varepsilon_x(0)}^{\varepsilon_x}\,Y(\varepsilon_x)\,d I(\varepsilon_x) 
\,=\,f^2\,X(\varepsilon_x)\,Y(\varepsilon_x)\,.
\end{equation} 
The equation (13), thus, defines the function $f\,=\,F(\varepsilon_x)$, 
and finally one gets $A\,=\,\varepsilon_{x}^{-1}\,F(\varepsilon_x)$.

Solving the equation (8) with respect to $f^{\prime}$ and taking into 
account (13), we find the equation
$$f^{\prime}\,=\,\frac{d F}{d \varepsilon_x}\,\varepsilon_{x}^{\prime}\,= 
\,\pm\, 
k\,K(\varepsilon_x)\biggl[I(\varepsilon_x)\,-\,\frac{F^2(\varepsilon_x)}
{\varepsilon_{x}^{2}}\biggr]^{1/2}\,,$$ which is integrated in 
quadratures:  \begin{equation} k\,x\,=\,\pm\,\int\limits_{}^{}\,\frac{d 
F(\varepsilon_x)} 
{K(\varepsilon_x)\biggl[I(\varepsilon_x)\,-\,\frac{F^2(\varepsilon_x)}
{\varepsilon_{x}^{2}}\biggr]^{1/2}}\,.
\end{equation}
Here the sign $\pm$ is to be chosen in accordance with the domain 
$x\,>\,0$ or $x\,<\,0$.  The integration constants in (14) are to be found 
from the boundary conditions, which are equivalent to the continuity 
conditions for the functions $f$ and $f^{\prime}$ at $x\,=\,0$. The 
solution of the boundary conditions, mentioned, is defined by the form of 
the functions $\varepsilon_x(|{\bf E}|^2)$ and $\varepsilon_z(|{\bf 
E}|^2)$.  

Let us consider a simple example assuming that
\begin{equation}
\varepsilon_x\,=\,a_x\,+\,b_x| {\bf E}|^2,\quad
\varepsilon_y\,=\,a_y\,+\,b_y| {\bf E}|^2, \quad
\varepsilon_z\,=\,a_z\,+\,b_z| {\bf E}|^2,
\end{equation}
with $a_x,\, a_y,\, a_z,\, b_x,\, b_y,\, b_z$ being constants.
Then from (7) one gets the functions:
\begin{equation}
I(\varepsilon_x)\,=\,\frac{\varepsilon_x\,-\,a_x}{b_x}, \quad
K(\varepsilon_x)\,=\,q\,+\,\frac{b_z}{b_x}\,\varepsilon_x,
\end{equation}
with $q\,=\,a_z\,-\,(b_z/b_x)\,a_x$.
Putting them into (11) we get (see Appendix 2)
\begin{equation}
Y(\varepsilon_x)\,=\,\bigl(q\,+\,\frac{b_z}{b_x}\,\varepsilon_x\bigr)\,
\biggl[\frac{2\,\beta^2\,\varepsilon_x\,+\,d\,-\,\sqrt{D}}
{2\,\beta^2\,\varepsilon_x\,+\,d\,+\,\sqrt{D}}\biggr]^
{d/b_x\,\sqrt{D}},
\end{equation}
where $d\,=\,b_z/b_x\,-\,1$ and
$D\,=\,(1\,+\,b_z/b_x)^2\,+\,4\,\beta^2\,q$. Here we assume that 
$D\,>\,0$. For simplicity we will now consider the case when the 
dielectric is isotropic, i.e. 
\begin{equation}
\varepsilon_x\,=\,\varepsilon_y\,=\,\varepsilon_z\,=\, 
\varepsilon\,=\,a\,+\,b\,| {\bf E}|^2.
\end{equation}
In this case one gets the integrating 
factor $Y(\varepsilon)\,=\,\varepsilon.$ Putting it into (13) and taking 
into account that at $x\,\to \infty, \quad f \to 0$ and 
$\varepsilon \to a$, we find 
\begin{equation}
f\,=\,F(\varepsilon)\,=\,\sqrt{\frac{\varepsilon(\varepsilon^2\,-\,a^2)} 
{2\,b\,(2\,-\,\beta^2\,\varepsilon)}}.
\end{equation}
The equation (14) in this case reads:
\begin{equation}
k\,x\,=\, \pm \,
\int\limits_{}^{}\frac{3\,\varepsilon^2\,-\,\beta^2\,\varepsilon^3 
\,-\,a^2}{\varepsilon\,(\varepsilon\,-\,a)\,(2\,-\,\beta^2\,\varepsilon)
\sqrt{P(\varepsilon)}}\,d\varepsilon,
\end{equation}
where 
$P(\varepsilon)\,=\,(\varepsilon\,+\,a)\,(3\,      
\varepsilon\,-\,2\,\beta^2\,\varepsilon^2\,-\,a).$ One can rewrite this 
equality in the form:
\begin{eqnarray}
k\,x\,&=&\,\pm\biggl[\int \frac{d\varepsilon}{\sqrt{P(\varepsilon)}}
\,+\,a\int 
\frac{d\varepsilon}{(\varepsilon - a)\sqrt{P(\varepsilon)}}\,+\, \nonumber 
\\ \,&+&\,\frac{a}{2}\int 
\frac{d\varepsilon}{\varepsilon\,\sqrt{P(\varepsilon)}} 
\,+\,(1+\frac{a\beta^2}{2})\int\frac{d\varepsilon}{(2-\beta^2
\varepsilon)\,\sqrt{P(\varepsilon)}}\biggr].
\end{eqnarray}
Using the substitution
\begin{equation}
\varepsilon\,=\,\frac{3}{4\,\beta^2}\,+\,\frac{\mu}
{4\,\beta^2}\,\mbox{cos}2\varphi, \quad 
\mu\,=\,\sqrt{9\,-\,8\,\nu}, \quad \nu\,=\,a\,\beta^2,   
\end{equation}
one can rewrite the last equality by means of elliptical integrals (see 
Appendix 3):  
\begin{equation}
k\,x\,=\,\mp\bigr[C_0\,F(\varphi,\, 
s)\,+\,\sum_{j=1}^{3}\,C_j\,\Pi_j (\varphi,\, n_j,\, s)\bigr],
\end{equation} 
where 
\begin{eqnarray}
C_0&=&\frac{2\sqrt{2}}{\xi}, \quad \qquad \qquad s^2=\frac{2\,\mu}{\xi^2},
\nonumber \\
C_1&=&\frac{8\sqrt{2}\nu}{\xi\,
(3-4\nu+\mu)},  \quad  n_1=-\frac{2\,\mu}{3-4\nu+\mu},
\nonumber \\
C_2&=&\frac{4\sqrt{2}\nu}{\xi\, 
(3+\mu)},   \qquad \quad n_2=-\frac{2\,\mu}{3+\mu},
\nonumber \\
C_3&=&\frac{8\sqrt{2}(1+\nu/2)}{\xi\,
(5+\mu)},  \quad n_3=-\frac{2\mu}{5+\mu},
\nonumber
\end{eqnarray}
with $\xi\,=\,\sqrt{3+4\nu+\mu}$.

The boundary conditions at $x\,=\,0$
\begin{equation}
\varepsilon_1\,E_{x1}\,=\,\varepsilon_2\,E_{x2},
\end{equation}
and 
\begin{equation}
E_{z1}\,=\,E_{z2}
\end{equation}
for the case considered can be represented as follows:
\begin{equation}
\frac{b_2}{b_1}\,\frac{\varepsilon_1\,(\varepsilon_1\,-\,a_1)
\,(\varepsilon_1\,+\,a_1)}{\beta_1^2\,\varepsilon_1\,-\,2}\,=\,
\frac{\varepsilon_2\,(\varepsilon_2\,-\,a_2)
\,(\varepsilon_2\,+\,a_2)}{\beta_2^2\,\varepsilon_2\,-\,2},
\end{equation}
\begin{equation}
\frac{(\varepsilon_1\,-\,q_1\,+\,\tau_1)
(\varepsilon_1\,-\,q_1\,-\,\tau_1)}
{\varepsilon_1^2\,(\varepsilon_1\,+\,a_1)}\,=\,
\frac{(\varepsilon_2\,-\,q_2\,+\,\tau_2)
(\varepsilon_2\,-\,q_2\,-\,\tau_2)}
{\varepsilon_2^2\,(\varepsilon_2\,+\,a_2)},
\end{equation}
where $q_i\,=\,3/4\beta_i^2$ and
$\tau_i\,=\,\sqrt{9-8a_i\beta_i^2}/4\beta_i^2, \quad i\,=\,1,2$.

The explicit form of relations (26) and (27) makes evident the existence 
of such parameters $b_2/b_1$ or $\beta$ for which the equations 
(26) and (27) are satisfied.  
\vskip 1cm 
\underline {Appendix 1}:\\

Multiplying the equation (10) by $Y$ one gets:
$$
Y\,\frac{d I}{d \varepsilon_x}\,\varepsilon_{x}^{\prime} \,=\, 
(f^2)^{\prime} \biggl(\frac{1}{\varepsilon_{x}^{2}}\,-\, 
\frac{\beta^2}{K}\,+\, 
\frac{1}{\varepsilon_x\,K}\biggr)\,Y\,-\,
\frac{2\,f^2}{\varepsilon_{x}^{3}}\, \varepsilon_{x}^{\prime}\,Y.$$
The righthand side of the last equation can be written as:
$$ \bigl[f^2\,X\,Y \bigr]^{\prime}
\,-\,\frac{2\,f^2}{\varepsilon_{x}^{3}}\, \varepsilon_{x}^{\prime}\,Y
\,-\,f^2\,X\,Y^{\prime}\,-\,f^2\,X^{\prime}\,Y.$$
Here $X\,=\,1/\varepsilon_{x}^{2}\,-\,\beta^2/K\,
+\,1/\varepsilon_x\,K.$
Equating the last three terms of the R.H.S. to zero we come to the equation
$$\frac{Y^{\prime}}{Y}\,=\,\frac{K^{\prime}}{X\,K^2}\,\bigl(\frac{1}
{\varepsilon_x}\,-\,\beta^2\bigr)\,+\,\frac{\varepsilon^{\prime}}
{\varepsilon^2\,K\,X},$$
which leads to the equation (11).

\vskip 1cm
\underline {Appendix 2}:\\

Putting $K(\varepsilon_x)$ and $X$ into (11) one gets
$$ \mbox{ln}Y(\varepsilon_x)\,=\,(b_z/b_x)\,\int\limits 
\frac{d\varepsilon_x}{K}\,-\,d\,\int\limits
\frac{d\varepsilon_x}{K\,+\,\Omega},$$
with $\Omega\,=\,\varepsilon_x\,(1\,-\,\beta^2\,\varepsilon_x)$.
Assuming that $D\,=\,(1\,+\,b_z/b_x)^2\,+\,4\,\beta^2\,q\,>\,0$ after 
integrating one finds:
$$\mbox{ln}Y(\varepsilon_x)\,=\,
\mbox{ln}\mid q\,+\,\frac{b_z}{b_x}\,\varepsilon_x\mid\,+\,
[d/b_x\,\sqrt{D}]\,\mbox{ln}
\mid\frac{2\,\beta^2\,\varepsilon_x\,+\,d\,-\,\sqrt{D}}
{2\,\beta^2\,\varepsilon_x\,+\,d\,+\,\sqrt{D}}\mid
.$$
\vskip 1cm
\underline {Appendix 3}:\\

For 
$\varepsilon\,=\,\frac{3}{4\,\beta^2}\,+\,\frac{\mu}
{4\,\beta^2}\,\mbox{cos}2\varphi,$
one gets
$ d\varepsilon\,=\,-\,\frac{\mu}{2\,\beta^2}\,
\mbox{sin}2\varphi\,d\varphi,$ and
$$P(\varepsilon)\,=\,\frac{\mu^2}{8\,\beta^2}\,\mbox{sin}^2 2\varphi
\,\biggl[\frac{(3\,+\,4a\beta^2)\,+\,\mu\,\mbox{cos}
2\varphi}{4\,\beta^2}\biggr].$$
In view of these expressions we get
$$\int\frac{d\varepsilon}{\sqrt{P(\varepsilon)}}\,=\,
-\,C_0\,\int\frac{d\varphi}{\sqrt{1\,-\,s^2\,\mbox{sin}^2\varphi}}\,=\,
-\,C_0\,F(\varphi,\,s).$$
Analogously one can find the other terms.
 \end{document}